# Field dependence of magnetization reversal by spin transfer

J. Grollier[1], V. Cros[1], H. Jaffrès[1], A. Hamzic[1,a], J.M. George[1], G. Faini[2], J. Ben Youssef[3], H. Le Gall[3], A. Fert[1]

[1] Unité Mixte de Physique CNRS-THALES and Université Paris-Sud, 91404 Orsay, France
[2] Laboratoire de Photonique et de Nanostructures, Route de Nozay, 91460 Marcoussis, France
[3] Laboratoire de Magnétisme de Bretagne, Université de Bretagne Occidentale, 29285 Brest, France
a) on leave from the Department of Physics, Faculty of Science, HR-10000 Zagreb, Croatia.

**Abstract**

We analyze the effect of the applied field ($H_{appl}$) on the current-driven magnetization reversal in pillar-shaped Co/Cu/Co trilayers, where we observe two different types of the transition between the parallel (P) and antiparallel (AP) magnetic configurations of the Co layers. If $H_{appl}$ is weaker than a rather small threshold value, the transitions between P and AP are irreversible and relatively sharp. For $H_{appl}$ exceeding the threshold value, the same transitions are progressive and reversible. We show that the criteria for the stability of the P and AP states and the experimentally observed behavior can be precisely accounted for by introducing the current-induced torque of the spin transfer models in a Landau-Lifshitz-Gilbert equation. This approach also provides a good description for the field dependence of the critical currents.

## I    Introduction

In 1996, Slonczewski[1] and also Berger[2] predicted that the magnetization of a magnetic layer can be reversed by the injection of a spin polarized current and the spin transfer to the layer. An experimental confirmation of the reversal of a magnetic moment without the application of an external magnetic field would have been of considerable interest for the magnetic switching of micro-devices, and these theoretical predictions prompted extensive experimental studies of the effect of spin polarized currents on the magnetic nanostructures[3-12]. The most quantitative results have been obtained on multilayered pillars[7, 9-11], typically Co/Cu/Co trilayers, in which the magnetic moment of a thin Co layer is reversed by the spin-polarized current injected from a thicker Co layer. These experiments have confirmed some of the main features predicted by the theory: (i) the effects induced by opposite currents are opposite: if the current of a given sign favors the parallel (P) magnetic configuration of the trilayer, the current of the opposite sign favors the antiparallel (AP) configuration; (ii) the current densities needed to switch such magnetic configuration are of the order of magnitude predicted by theory i.e. $10^7$ A/cm$^2$. On the other hand, the experimental data have not yet established clearly the variation of the critical currents with the layer thickness, nor has the effect of an applied magnetic field been fully understood.

From the theoretical point of view, several models have been developed. Most of them[13-20] are based on Slonczewski's concept[1] of spin transfer involving the transverse components of the current spin polarization. Another approach, proposed by Heide[21], involves the longitudinal components of the polarization and the effect of the current is expressed by an effective exchange-like interaction between the magnetic moments of the two magnetic layers.

Here we focus on the dependence of the magnetization reversal induced by a spin current in the presence of an applied field $H_{appl}$. We will analyze our experimental results on Co/Cu/Co pillars (that were partially published elsewhere[9]) and show the existence of two different field regimes. If the applied fields do not exceed a fairly small threshold value, irreversible and relatively sharp transitions between parallel (P) and antiparallel (AP) magnetic configurations of the Co layers take place. In a second regime, for fields above this threshold value, the transition between P and AP is progressive and reversible. We will explain that this behavior can be well accounted for by introducing the current-induced torque of the spin transfer models in a Landau-Lifshitz-Gilbert equation to study the stability of the P and AP states. It will also be shown that the existence of these different regimes of the field dependence of the critical currents cannot be justified with the model describing the effect of the current as an effective exchange-like interaction between the magnetic moments of the cobalt layers[21].

In Section II, we present the first set of our experimental results, which is the current dependence of the pillar resistance measured for different (constant) applied magnetic fields. These data give clear evidence for the two very different types of behavior. In Section III, we give the theoretical analysis and put forward the criteria for the stability of the P and AP configurations. Section IV is devoted to a different experimental approach for the study the magnetization reversal by spin transfer, that is the field dependence of the pillar resistance measured with constant currents. The field dependence of the critical currents is analyzed in Section V. Finally, the details of our calculations are given in Appendix I (for $H_{appl} > 0$) and Appendix II (for $H_{appl} < 0$).

## II  Resistance *vs*. current at constant applied magnetic field

Our experiments have been performed on submicronic (200 x 600 nm$^2$) pillars, (fabricated by e-beam lithography), which central part is a Co1(15nm)/Cu(10nm)/Co2(2.5nm) trilayer. Details on the fabrication have been described elsewhere[9]. The trilayer exhibits CPP-GMR effects, with a difference of about 1mΩ between the resistances in the P and AP configurations. This change of the resistance (GMR effect) has been used to determine the changes of the magnetic configuration of the trilayer.

For all the experimental data presented here, the initial magnetic configuration (prior to the injection of a DC current) has been a parallel (P) magnetic configuration of the system, in which the magnetic moments of the Co layers are along the positive direction of an axis parallel to the long side of the rectangular pillar, while the applied magnetic field $H_{appl}$ (either zero or a constant positive value) is applied along the positive direction of the same axis (thus stabilizing this initial P magnetic configuration). An increasing or decreasing DC current (*I*) is then applied, and the variation of the pillar resistance (*R*) with this current is recorded. The results reported here were obtained at 30 K (we point out that the critical currents are smaller at room temperature). In our convention, a positive DC current corresponds to the electron flow from the thick Co layer towards the thin one.

In Fig. 1, we present the variation of the resistance $R$ as a function of the DC current $I$ for $H_{appl} = 0$ and 125 Oe. Starting from a P configuration (for $I = 0$) and increasing the current towards the positive values, only a progressive and reversible small increase of the resistance $R$ can be observed, which can be ascribed to the heating of the sample (this has been also seen in all other experiments on pillars[7, 9-11], when current densities reach the value of $10^7$ A/cm$^2$). In contrast, when the current is negative and at a critical value $I_C^{P \to AP}$, an irreversible jump of the resistance ($\Delta R \approx 1$ m$\Omega$) is clearly seen, which corresponds to a transition from the P to the AP configurations, thus indicating the reversal of the magnetic moment of the thin Co layer. The trilayer then remains in this high resistance state (the $R_{AP}(I)$ curve) until the current is swept to positive values, where, at the critical current $I_C^{AP \to P}$, the resistance drops back to the $R_P(I)$ curve. In a small range of the applied magnetic field (that we shall note as the *regime A*), this type of hysteretic $R(I)$ curve is the fingerprint of the magnetization reversal by spin injection[7,9-11].

If the applied magnetic field is zero, $I_C^{P \to AP} \cong -19$ mA (corresponding to the current density $j_C^{P \to AP} \cong -1.58 \cdot 10^7$ A/cm$^2$) and $I_C^{AP \to P} \cong + 14$ mA ($j_C^{AP \to P} \cong -1.17 \cdot 10^7$ A/cm$^2$). For positive magnetic fields, which stabilize the P configuration, $|I_C^{P \to AP}|$ increases and $I_C^{AP \to P}$ decreases. This is seen in Fig. 1, if one compares the critical currents values obtained for $H_{appl} = 0$ and $H_{appl} = + 125$ Oe. We emphasize, however, that the shift of $I_C^{AP \to P}$, induced by the applied field, is larger than that of $I_C^{AP \to P}$.

The $R(I)$ curve for $H_{appl} = + 500$ Oe (shown in Fig. 2) illustrates the different behavior, observed when the applied fields are higher (and we shall note this as the *regime B*). Starting from $I = 0$ in a P configuration (on the $R_P(I)$ curve), a large enough negative current still induces a transition from P to AP, but now this transition is very progressive and reversible. The $R(I)$ curve departs from the $R_P(I)$ curve at $I_{start}^{P \to AP} \cong -25$ mA ($j_{start}^{P \to AP} \cong -2.08 \cdot 10^7$ A/cm$^2$), reaches finally $R_{AP}(I)$ at $I_{end}^{P \to AP} \cong -45$ mA ($j_{end}^{P \to AP} \cong -3.75 \cdot 10^7$ A/cm$^2$) and for the higher negative values, the resistance continues to follow the $R_{AP}(I)$ curve. On the way back (towards the positive current values) $R(I)$ departs from $R_{AP}(I)$ at $I_{start}^{AP \to P} \equiv I_{end}^{P \to AP} \cong -45$ mA and reaches finally $R_P(I)$ at $I_{end}^{AP \to P} \equiv I_{start}^{P \to AP} \cong -25$ mA. If the same type of experiment is done at even higher values of the applied field, the progressive and reversible transitions have the same behavior, but they occur at higher negative current values. Finally, for very large applied field ($H_{appl} = 5000$ Oe), the transition is out of our experimental current range, and the recorded curve is simply $R_P(I)$.

## III The calculation of the critical currents in the presence of an external field.

In order to study the stability or instability of a P (AP) configuration in presence of a DC current, we will treat the motion of the magnetic moment of the thin cobalt layer by using the Landau-Lifshitz-Gilbert (LLG) equation, in which we will introduce a current-induced torque of the form predicted by Slonczewski[1]. This approach is certainly less quantitatively precise than those based on micromagnetics simulations[22] with a non-uniform magnetization, but, as we will see, it can nevertheless account for most of the qualitative features of the experimental results.

We shall denote the unit vectors along the magnetic moments of the thick and thin Co layers as **m₁** and **m₂** respectively. We suppose that there is an uniaxial magnetic anisotropy in the layer plane along the x axis (the long side of the rectangular layers in our experiments) and that **m₁** is fixed in the positive direction of this axis. We put $\mathbf{u_x} \equiv \mathbf{m_1}$ for the unit vector along x axis, and $\mathbf{u_z}$ for the unit vector along the z axis, which is perpendicular to the layers. The magnetic field $H_{appl}$ is (as in our experiments) applied along the x axis. The stability conditions for the P or AP configurations are obtained by studying the motion of **m₂**, when the angle θ between **m₂** and $\mathbf{u_x}$ is close to either 0 or π. The LLG equation can be written as:

$$\frac{d\vec{m}_2}{dt} = -\gamma_0 \vec{m}_2 \times [H_{eff} \vec{u}_x - H_d (\vec{m}_2 \cdot \vec{u}_Z) \vec{u}_Z]$$
$$+ \alpha \vec{m}_2 \times \frac{d\vec{m}_2}{dt} - G^{P(AP)} j\, \vec{m}_2 \times (\vec{m}_2 \times \vec{u}_x)$$
(1)

where
$$H_{eff} = H_{appl} \pm H_{an} \quad (2)$$

$H_d = 4\pi M_s$ describes the out-of-plane anisotropy induced by the demagnetizing field, $H_{an}$ is the in-plane uniaxial anisotropy, + or - depend whether the configuration is close to P or AP, and $\alpha$ is the Gilbert damping coefficient. The last term in Eq.(1) is the contribution from the spin torque[1], where $j$ is the current density and

$$G^{P(AP)} = \frac{2\mu_B P_S^{P(AP)}}{t_2 M_s e} \quad (3)$$

The coefficient $P_S$ is the spin polarization and takes different value ($P^P$ or $P^{AP}$) depending on whether **m₂** is close to either the P or the AP configuration, $t_2$ is the thickness of the thin Co layer and $M_s$ is the Co magnetization.

At this point, it could be noted that, in the analysis of their experimental results, Katine *et al*[7] have used the equations similar to our Eq.(1), to calculate the critical currents. However, they have considered one particular case only. They first solved the LLG equation without the Gilbert and current-induced terms and derived the equations for the small periodic elliptical precession of **m₂** around $\mathbf{u_x}$ (or $-\mathbf{u_x}$), generated by only the field terms of Eq.(1). They then calculated the work of the Gilbert and current-induced terms of Eq.(1) (during a period of this elliptical precession), and deduced the stability or instability of the configuration from the sign of the calculated work. This method can be applied only when the motion, generated by the applied, anisotropy and demagnetizing field, is a periodic precession; i.e. for $H_{appl} < H_{an}$ (when **m₂** is close to $-\mathbf{u_x}$; AP configuration) and for $H_{appl} > -H_{an}$ (when **m₂** is close to $+\mathbf{u_x}$; P configuration). Our present calculation is more general and holds for any value of the applied field. As a consequence, we will show that the *regime B* of our experimental results is expected in a field range where the calculation made by Katine *et al*[7] cannot be applied.

Projecting the LLG equations onto the three axes x, y and z, we obtain the following equations for the components $m_x$, $m_y$ and $m_z$ of **m₂**:

$$\dot{m}_x = \gamma_0 H_d m_z m_y + \alpha\, m_y \dot{m}_z - \alpha\, m_z \dot{m}_y$$
$$+ G^{P(AP)} j (m_y^2 + m_z^2)$$

$$\dot{m}_y = -\gamma_0 H_{eff} m_z - \gamma_0 H_d m_z m_x + \alpha\, m_z \dot{m}_x$$
$$- \alpha\, m_x \dot{m}_z - G^{P(AP)} j m_x m_y \quad (4)$$

$$\dot{m}_z = \gamma_0 H_{eff} m_y + \alpha\, m_x \dot{m}_y - \alpha\, m_y \dot{m}_x$$
$$- G^{P(AP)} j m_x m_z$$

When the angle θ between $\mathbf{m_2}$ and $\mathbf{u_x} = \mathbf{m_1}$ is small (or close to π), by keeping only the terms of first order in $m_y$ and $m_z$ and also neglecting the terms in $\alpha^2$ (the Gilbert coefficient is a small number), Eq.(4) can be written as

$$m_x = \pm 1$$
$$\dot{m}_y = \mp(\alpha\alpha_0 H_{eff} + Gj) m_y$$
$$\quad + [-\gamma_0(H_{eff} \pm H_d) + \alpha Gj] m_z$$
$$\dot{m}_z = (\gamma_0 H_{eff} - Gj) m_y$$
$$\quad + [-\alpha\gamma_0(\pm H_{eff} + H_d) \mp \alpha Gj] m_z$$
$$(5)$$

where ± (∓) means + (−) when the configuration is close to P, and − (+) when the configuration is close to AP. Also, $G$ is $G^P$ or $G^{AP}$. The general solutions for $m_y$ and $m_z$ are of the form:

$$A\exp(k_1 t) + B\exp(k_2 t) \quad (6)$$

The condition for the instability of a given magnetic configuration (P or AP) is related to the sign of the real part of $k_1$ and $k_2$: a positive sign means that the amplitude of the motion of $\mathbf{m_2}$ increases with time (and the instability occurs). $k_1$ and $k_2$ are the solutions of the quadratic equation which, after dropping the terms of second order in α, is written as

$$k^2 \pm 2k[2\gamma_0(H_{eff} \pm \frac{H_d}{2}) \pm Gj] + G^2 j^2$$
$$+ \gamma_0^2 H_{eff}(H_{eff} \pm H_d) = 0$$
$$(7)$$

The solution of Eq.(7) and the expressions for $k_1$ and $k_2$ are detailed in Appendix I. Here, we focus on the results corresponding to our experiments, i.e. when $H_{appl}$ is positive and thus favors the orientation of $\mathbf{m_2}$ (thin layer) in the direction parallel to $\mathbf{u_x} = \mathbf{m_1}$ (thick layer). The results for negative values of $H_{appl}$ are presented in Appendix II.

The overall behavior for $H_{appl} > 0$ can be separated into three different regimes, which we will now discuss separately. They are also schematically depicted in Fig.3.

*(A)*     $H_{appl} > 0$ and $H_{an} - H_{appl} \gg \alpha^2 H_d$

This regime occurs when $|H_{an}-H_{appl}|$ is larger than few tens of Oe, (if one takes the α value of Co derived from FMR[23]). The P configuration becomes unstable when the sign of the real part of $k_1$ and $k_2$ is positive, that is for

$$j < -\frac{\alpha\gamma_0}{G^P}(H_{appl} + H_{an} + \frac{H_d}{2})$$

whereas the AP configuration is unstable for

$$j > +\frac{\alpha\gamma_0}{G^{AP}}(-H_{appl} + H_{an} + \frac{H_d}{2}).$$

Thus, starting from a P configuration at zero current, the increase of the current to the negative values results in an unstable P configuration, which can switch directly to a stable AP configuration at the critical current density $j_C^{P \to AP}$ (this corresponds to the point M in Fig.3.):

$$j_c^{P \to AP} = -\frac{\alpha\gamma_0}{G^P}(H_{appl} + H_{an} + \frac{H_d}{2})$$
(8)

When the current returns to zero and becomes positive, the AP configuration becomes unstable, and it can switch directly to a stable P configuration at the critical current density $j_c^{AP \to P}$ (point N in Fig.3):

$$j_c^{AP \to P} = +\frac{\alpha\gamma_0}{G^{AP}}(-H_{appl} + H_{an} + \frac{H_d}{2})$$
(9)

Such a hysteretic behavior, with direct transitions between P and AP, has been predicted by Katine *et al*[7], and it also corresponds to our experimental observations (cf. Fig.1.).

**(B)** $H_{appl} > 0$ and $H_{appl}-H_{an} \gg \alpha^2 H_d$

The condition of instability of the P configuration (positive signs of $k_1$ and $k_2$) is similar to the one derived for the *case A*, i.e.

$$j < -\frac{\alpha\gamma_0}{G^P}(H_{appl} + H_{an} + \frac{H_d}{2})$$

On the other hand, the condition for an stable AP configuration has changed and becomes

$$j < -\frac{\gamma_0}{G^{AP}}[(H_{appl} - H_{an})(H_{appl} - H_{an} + \frac{H_d}{2})]^{\frac{1}{2}}$$

In this case, starting from a P configuration at zero current, the increase of the current to the negative values results in an unstable P configuration at the critical current density $j_{start}^{P \to AP}$ (point Q in Fig.3):

$$j_{start}^{P \to AP} = -\frac{\alpha\gamma_0}{G_P}(H_{appl} + H_{an} + \frac{H_d}{2}) \quad (10)$$

However, the AP configuration is not stable at this current density value. Taking into account the condition for the stability around $\theta = \pi$, the AP configuration is reached only at the critical current density $j_{end}^{P \to AP}$ (point R in Fig.3):

$$j_{end}^{P \to AP} = -\frac{\gamma_0}{G^{AP}}[(H_{appl} - H_{an})(H_{appl} - H_{an} + H_d)]^{\frac{1}{2}}$$
(11)

A straightforward check of Eq. (10) confirms that, for $(H_{appl} - H_{an}) \gg \alpha^2 H_d$ and $H_d \gg H_{an}$, $|j_{end}^{P \to AP}| > |j_{start}^{P \to AP}|$.

When the current is swept back, the AP configuration becomes unstable at the same critical current density (cf. Eq.(11) and point R in Fig.3):

$$j_{start}^{AP \to P} = j_{end}^{P \to AP} \quad (12)$$

The P configuration is reached only for:

$$j_{end}^{AP \to P} = j_{start}^{P \to AP} \quad (13)$$

which is the point Q in Fig. 3.

**(C)** $H_{appl} > 0$ and $H_{appl}-H_{an} \approx \alpha^2 H_d$

The condition for an unstable P configuration is the same as in case A or B. On the other hand, when $\theta$ is close to $\pi$, there is no simple analytical solution of Eq.(5) if $H_{appl}$ is in a zone of width $\approx \alpha^2 H_d$ around $H_{an}$ (cf. Fig.3). The dotted line in Fig. 3 is what is qualitatively expected for the variation of the critical current of the AP configuration in the crossover zone.

In Fig. 3, we draw also the main features of the expected *R(I)* dependences for zero and non-zero values of the applied field, shown in the insets. The insets (α) and (β) represent the expected *R(I)* variations in the *regime A*. This behavior is comparable to what is observed experimentally at low field (Fig. 1). In Eqs.(8-9), the asymmetry between the currents, for the P→AP and AP→P transitions (inset α), comes from the difference between $G^P$ and $G^{AP}$. When $H_{appl}$ increases (inset β), both transitions are shifted to the left. (The observed larger shift for $j_c^{AP \to P}$ in Fig.1 is probably due to the deviations from Eq.(9) when one approaches the crossover region between *regimes A* and *B*). The inset (γ) represents the *R(I)* curve expected in the *regime B*. The slope corresponds to the progressive and reversible transition between Q and R in Fig.3. A similar behavior is observed experimentally at high field (cf. Fig. 2). We do emphasize again that such a characteristic cannot be predicted by a calculation which assumes that the motion of **m₂** is a periodic precession (as it was done by Katine *et al*[7]), which is obviously not the case in *regime B*.

In order to extend our approach to a more quantitative level, we have also calculated the critical currents in both regimes and predicted by Eqs. (8-11). The parameter used were: $t_2 = 2.5$ nm, $P^P = 0.07$ and $P^{AP} = 0.41$ (derived in the model of Fert *et al*[20] from CPP-GMR data on Co/Cu multilayers[24-25]), α = 0.007 (Ref. 23), $H_d$ = 1.79 T and $H_{an}$ = 150 Oe, which is approximately the field of the crossover between *regimes A* and *B* in our experiments (this value is also close to the value of $H_{an}$ derived from the numerical calculations of Chen *et al.*[26] for rectangular prisms).

In *regime A* and in zero applied magnetic field, we obtain $j_c^{P \to AP}$ = -4.9·10⁷A/cm² (experimentally: -1.6·10⁷A/cm²) and $j_c^{AP \to P}$ = +0.8·10⁷A/cm² (exp.: +1.17·10⁷A/cm²). For $H_{appl}$ = 500 Oe in the regime B, we obtain $j_{start}^{P \to AP}$ = -5.1·10⁷A/cm² (exp.: -1.7·10⁷A/cm²) and $j_{end}^{P \to AP}$ = -33·10⁷A/cm² (exp.: -4.25·10⁷A/cm²). This shows that the expressions of the spin transfer model predict the right order of magnitude for the critical currents in both *A* and *B regimes*. The stronger discrepancy for $j_{end}^{P \to AP}$ may be due to the difficulty in the precise determination of the point where the R(I) curve merges into the $R_{AP}(I)$ one (cf. Fig.2) which could result in the underestimated value of $j_{end}^{P \to AP}$.

Finally, it is interesting to see what are the conditions for the occurrence of the instabilities, if the effect of the current is described by an effective interaction energy of the form $E_{int}$ = -gj**m₁.m₂**, as in the model proposed by Heide[21]. This interaction can be expressed by an effective field gj**u_x** which adds to $H_{eff}$**u_x** in the first term of the LLG equation (cf. Eq.(1)). Following the same lines of reasoning (as we have done so far) for the stability (or instability) of the P and AP configurations, one can predict only a hysteretic behavior with direct transitions for:

$$j_c^{P \to AP} = -(H_{appl} + H_{an})/g$$
$$\text{and} \quad j_c^{AP \to AP} = (H_{an} - H_{appl})/g \qquad (14)$$

In other words, this approach does not predict a non-hysteretic and reversible reversal, which is in clear contradiction with our experimental results at high field. Furthermore, (as we will also show later) the field dependence of the critical currents expected from Eq.(14) is not consistent with the experimental observations.

## IV Resistance *vs*. applied magnetic field at constant DC current

In Fig.4, we present the variation of the resistance (*R*) of a pillar as a function of the applied field ($H_{appl}$) for several values of the DC currents (*I* = + 50, - 30, -40 and -50 mA). This type of experiments has not been used extensively so far, although, as it will be shown, it can also be very useful.

The *R($H_{appl}$)* curve for *I* = +50 mA is flat, i.e. there is no GMR. This means that a large positive current is able to maintain the P configuration of the Co magnetic moments throughout our experimental field range. This is the case that corresponds to a trilayer with a strong ferromagnetic interlayer exchange.

For the negative currents, on the other hand, the *R($H_{appl}$)* dependences have a bell-shaped appearance, which mimics to the GMR curve of an antiferromagnetically coupled trilayer. This resemblance is further supported by the fact that these *R($H_{appl}$)* curves broaden when the current increases – the same feature exhibited by a typical GMR curve when the strength of the AF coupling increases. For example, at I = - 50 mA, starting from the high positive field, the P configuration becomes unstable at about $H_{appl}$ = + 5600 Oe, where the resistance starts increasing. In Fig.3, this corresponds to the point S, at the interception of the horizontal dashed line with the $j_{start}^{P \to AP}$ curve (in the *regime B* of the diagram, as it should be expected for high applied fields). As $H_{appl}$ decreases below 5600 Oe, the resistance progressively (and reversibly) increases towards its maximum value of the AP configuration. This again is in agreement with the progressive and reversible character of the transitions in the *regime B*. In accordance with our scheme, shown in Fig.3, a stable AP configuration (a maximum of the resistance) is reached at the point T (the interception of the dashed line with the $j_{end}^{P \to AP}$ curve). When $H_{appl}$ becomes negative, the moment $m_1$ of the thick Co layer is reversed and, in the presence of a large negative current at low fields (*regime A*), one expects that a sharp reversal of $m_2$ (restoring the AP configuration) follows immediately. This explains why there is practically no discontinuity of the *R($H_{appl}$)* curves in the region of their maximum.

Quantitative values of $j_{start}^{P \to AP}$ for a given applied field can be extracted from the *R($H_{appl}$)* curves, for example for I = -50 mA at 5600 Oe, and these data will be used in the field dependence of the $j_C^{P \to AP}$ in the next section. On the other hand, it is clear (from the low field part of the *R($H_{appl}$)* curves shown) that the value of the applied magnetic field, at which the high resistance state is reached for a given current, cannot be reliably estimated. This is why we will not discuss the field dependence of the critical currents for the transitions from AP to P configuration.

## V Field dependence of the critical currents

The last part of our discussion deals with the field dependence of the critical currents for which the instability of the P configuration occur. Here again, we have compared our experimental data with the theoretical predictions, and this is shown in Fig.5. The data for the critical current $j_{start}^{P \to AP}$ of *regime B* are taken from *R($H_{appl}$)* data similar to those of Fig.4. For the *regime A*, which is extremely narrow at the scale of the figure, we have plotted only the zero-field $j_c^{P \to AP}$ value. It can be seen that all our experimental points are on the same straight line.

According to Eqs.(8,10), both $j_c^{P \to AP}$ and $j_{start}^{P \to AP}$ are expected to vary as

$$j_c^{P \to AP}(H_{appl}=0)\left[1+\frac{H_{appl}}{H_{an}+H_d/2}\right].$$

This predicted variation with $H_d$ and $H_{an}$ is represented as a dotted line on Fig.5. The agreement between the slopes of the experimental and calculated lines is rather satisfactory, which further supports our present approach. The same experimental data could also be compared with what is expected when the effect of the current is described by an effective interaction[21] $\propto$ **m$_1$.m$_2$** (given by Eq.(14) and plotted as a dashed line in Fig.5). Again, this approach is in a strong disagreement with the experimental field dependence of the critical currents.

## VI Conclusions

The following main conclusions can be derived from our experimental results and their analysis.

1) The experimental results for the magnetization reversal by spin transfer in the presence of the finite external magnetic field show unambiguously the existence of two qualitatively different regimes: *a low field regime A*, with direct and irreversible transitions between the P and AP configurations of the trilayer, and *a high field regime B* with progressive and reversible transitions. The intermediate range between $j_{start}$ and $j_{end}$ in *regime B* is supposed to correspond to a situation with current-maintained precession and spin wave emission[3,11].

2) The existence of the *regimes A* and *B* can be theoretically explained by a calculation in which Slonczewski's spin torque is introduced in the LLG equation to study the stability of the P and AP configurations. This approach enables one to establish a schematic diagram (Fig. 3) for the transition between these two regimes.

3) The field dependence of the critical currents can be reasonably well accounted for by the prediction derived from the LLG equations with Slonczewski's spin torque. The application of the model in which the effect of the current is expressed by an effective exchange-like interaction, disagrees with the experimental results.

**Appendix I**

When the trilayer is close to the P configuration (θ close to zero), the determinant of Eq.(7) is:

$$\Delta = 2\alpha\gamma_0 G^P j (H_{appl} + H_{an} + \frac{H_d}{2}) - \gamma_0^2 (H_{appl} + H_{an})(H_{appl} + H_{an} + H_d) \quad (A1)$$

A straightforward numerical estimate shows that, for $H_{appl} > 0$ and even for current densities largely exceeding the experimental range, $\Delta$ is negative. Consequently:

$$k_1 = -G^P j - \alpha\gamma_0 (H_{appl} + H_{an} + \frac{H_d}{2}) + i\sqrt{(-\Delta)}$$
$$k_2 = -G^P j - \alpha\gamma_0 (H_{appl} + H_{an} + \frac{H_d}{2}) - i\sqrt{(-\Delta)} \quad (A2)$$

The imaginary part of $k_1$ and $k_2$ is related to the ellipsoidal magnetization precession around the x axis due to the different applied magnetic fields. The real part of $k_1$ and $k_2$ describes the stability of the P state. For $j < -\frac{\alpha\gamma_0}{G^P}(H_{appl} + H_{an} + \frac{H_d}{2})$, $k_1$ and $k_2$ are positive, *exp(k₁t)* end *exp(k₂t)* increase with time, which means that the P state is unstable.

The same approach can be applied to discuss the stability of the AP configuration, but the problem is more complex. The determinant is now:

$$\Delta = -2\alpha\gamma_0 G^{AP} j (-H_{appl} + H_{an} + \frac{H_d}{2}) + \gamma_0^2 (H_{appl} - H_{an})(-H_{appl} + H_{an} + H_d) \quad (A3)$$

For $H_{appl} > 0$ and $H_{an}-H_{appl} \gg \alpha^2 H_d$, the first term in $\Delta$ can be neglected and $\Delta = \gamma_0^2 (H_{appl} - H_{an})(-H_{appl} + H_{an} + H_d)$ is negative. Thus,

$$k_1 = G^{AP} j - \alpha\gamma_0 (-H_{appl} + H_{an} + \frac{H_d}{2}) - i\sqrt{|\Delta|}$$
$$k_2 = G^{AP} j - \alpha\gamma_0 (-H_{appl} + H_{an} + \frac{H_d}{2}) + i\sqrt{|\Delta|} \quad (A4)$$

The real parts of k1 and k2 are positive and the AP configuration is unstable for :

$$j > \frac{\alpha\gamma_0}{G^{AP}}(\frac{H_d}{2} - H_{appl} + H_{an}) \quad (A5)$$

For $H_{appl} > 0$ and $H_{appl}-H_{an} \gg \alpha^2 H_d$, the first term in $\Delta$ can be neglected as well. $\Delta$ is positive and

$$k_1 = G^{AP} j - \alpha\gamma_0(-H_{appl} + H_{an} + \frac{H_d}{2}) - \sqrt{\Delta}$$
$$k_2 = G^{AP} j - \alpha\gamma_0(-H_{appl} + H_{an} + \frac{H_d}{2}) + \sqrt{\Delta}$$
(A6)

With $\sqrt{\Delta} \gg \alpha\gamma_0(-H_{appl} + H_{an} + \frac{H_d}{2})$, the AP state is unstable if $k_1$ or $k_2$ is positive. This leads to the condition for the instable AP state :

$$j > -\frac{\gamma_0}{G^{AP}} \sqrt{(H_{appl} - H_{an})(H_d - H_{appl} + H_{an})} \qquad (A7)$$

**Appendix II**

The calculation of critical currents for negative applied magnetic fields is similar to that developed in Section III and Appendix I. Nevertheless, the calculation is complicated by the thick layer reversal in negative fields. We will consider in this appendix that this reversal occurs at a field $H_{thick}$ smaller than the thin layer coercive field. The main results are:

In a low field regime where $|H_{appl}| < |H_{thick}|$ and $H_{an} - |H_{appl}| \gg \alpha^2 H_d$, there is a hysteretic behavior, similar to *regime A* for $H_{appl} > 0$. The direct transitions between P and AP states occur for:

$$j_c^{P \to AP} = -\frac{\alpha\gamma_0}{G^P}(H_{appl} + H_{an} + \frac{H_d}{2}) \qquad (A8)$$
$$j_c^{AP \to P} = +\frac{\alpha\gamma_0}{G^{AP}}(-H_{appl} + H_{an} + \frac{H_d}{2})$$

In another low field regime where $|H_{appl}| > |H_{thick}|$ and $H_{an} - |H_{appl}| \gg \alpha^2 H_d$, the thick layer has changed direction. The behavior is still hysteretic and the transitions between P and AP are given by :

$$j_c^{P \to AP} = -\frac{\alpha\gamma_0}{G^P}(-H_{appl} + H_{an} + \frac{H_d}{2})$$
$$j_c^{AP \to P} = +\frac{\alpha\gamma_0}{G^{AP}}(H_{appl} + H_{an} + \frac{H_d}{2}) \qquad (A9)$$

In a high field regime with the thick layer reversed towards the x < 0 direction and for $|H_{appl}| - H_{an} \gg \alpha^2 H_d$, there is a progressive and reversible transition between the P and AP configuration at:

$$j_{start}^{P \to AP} = -\frac{\alpha \gamma_0}{G_P}(H_{appl} + H_{an} + \frac{H_d}{2}) \quad\quad\quad (A10)$$

$$j_{end}^{P \to AP} = -\frac{\gamma_0}{G^{AP}}[|H_{appl} - H_{an}|(H_{appl} - H_{an} + H_d)]^{\frac{1}{2}}$$

**Figure captions**

**Figure 1:** Resistance *vs.* DC current for for sample 1 ($H_{appl} = 0$ (grey) and $H_{appl} = 125$ Oe (black)).

**Figure 2:** Resistance *vs.* DC current for for sample 2 ($H_{appl} = +500$ Oe (grey) and $H_{appl} = 0$ (black)).

**Figure 3:** Schematic representation of the critical currents *vs.* applied field deduced from the discussion of the stability of the P and AP configurations within the spin transfer model and LLG equation (in agreement with our approach, only the $H_{appl} > 0$ case is considered here). The solid curves correspond to the relevant equations (see text) for the critical currents in *regimes A* and *B*. The dotted curve in the zone $\approx \alpha^2 H_d$ is a guide for the eye. The evolution of the system along the vertical and horizontal dashed lines is described in the text. Insets indicate the expected R(I) curves: ($\alpha$) - *regime A,* zero applied field; ($\beta$) - *regime A,* non-zero applied field; ($\gamma$) *regime B*.

**Figure 4:** Resistance *vs.* applied magnetic field for sample 2 ($I_{DC}$ = - 50 mA (open squares), - 40 mA (solid squares), - 30 mA (open circles) and + 50 mA (solid circles)). For clarity, the curves have been shifted vertically to have the same high field baseline.

**Figure 5:** Field dependence of the critical current for the transition from P to AP ($j_c^{P \to AP}$ in *regime A* and $j_{start}^{P \to AP}$ in *regime B*). The symbols represent the experimental data for sample 2. The dotted line *S* is the expected variation in the spin transfer model and based on Eqs. (11, 13). The dashed line *H* is the expected variation in Heide's and model based on Eq. (14) (see text for details).

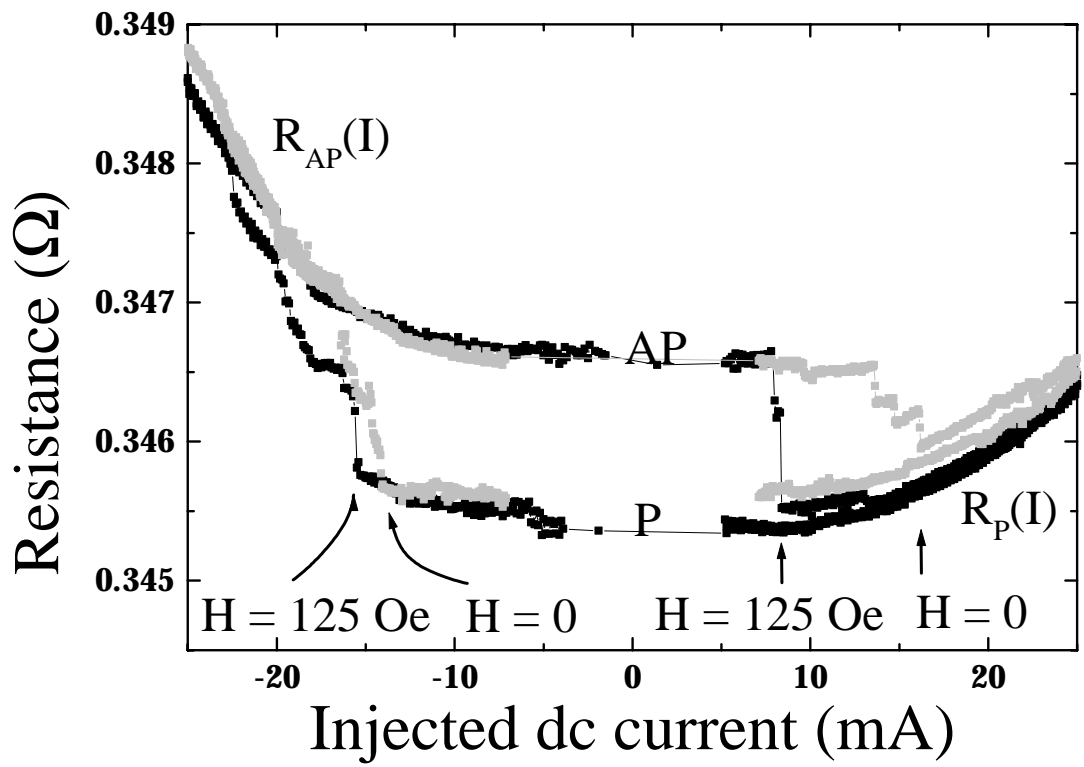

Figure 1

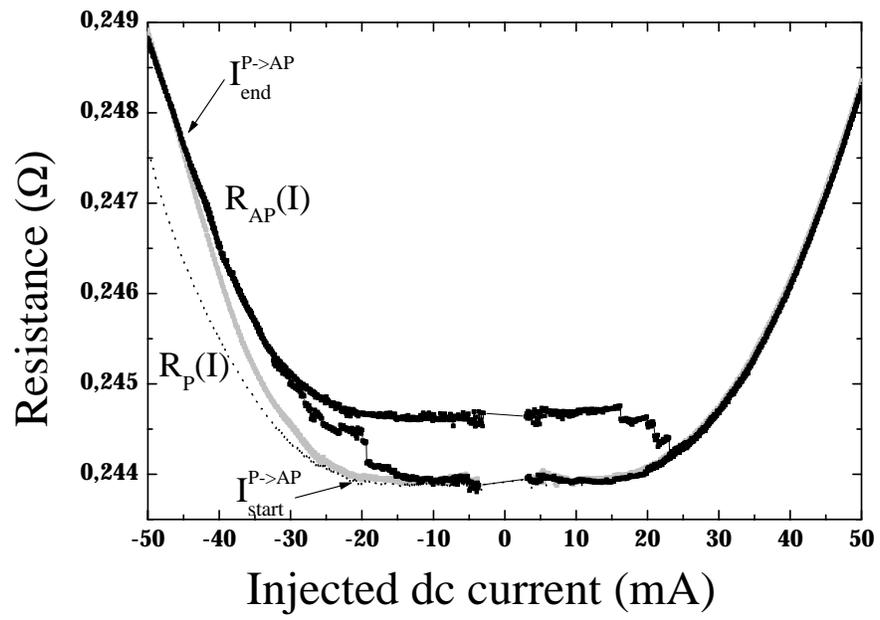

**Figure 2**

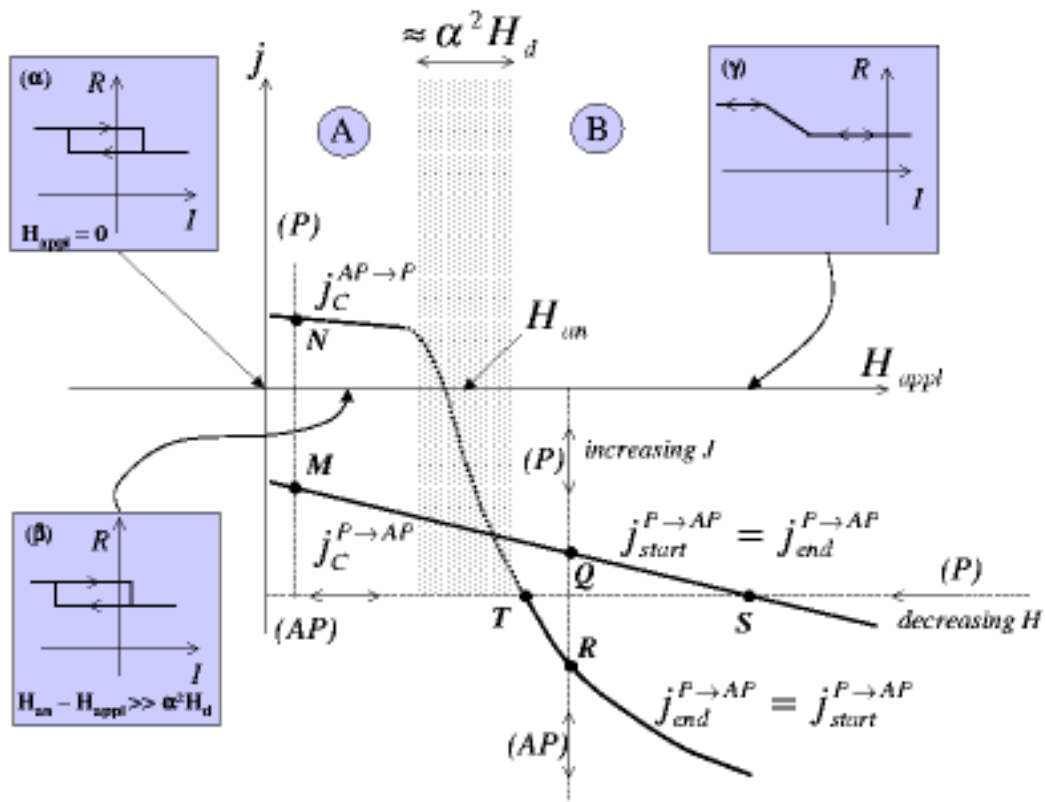

**Figure 3**

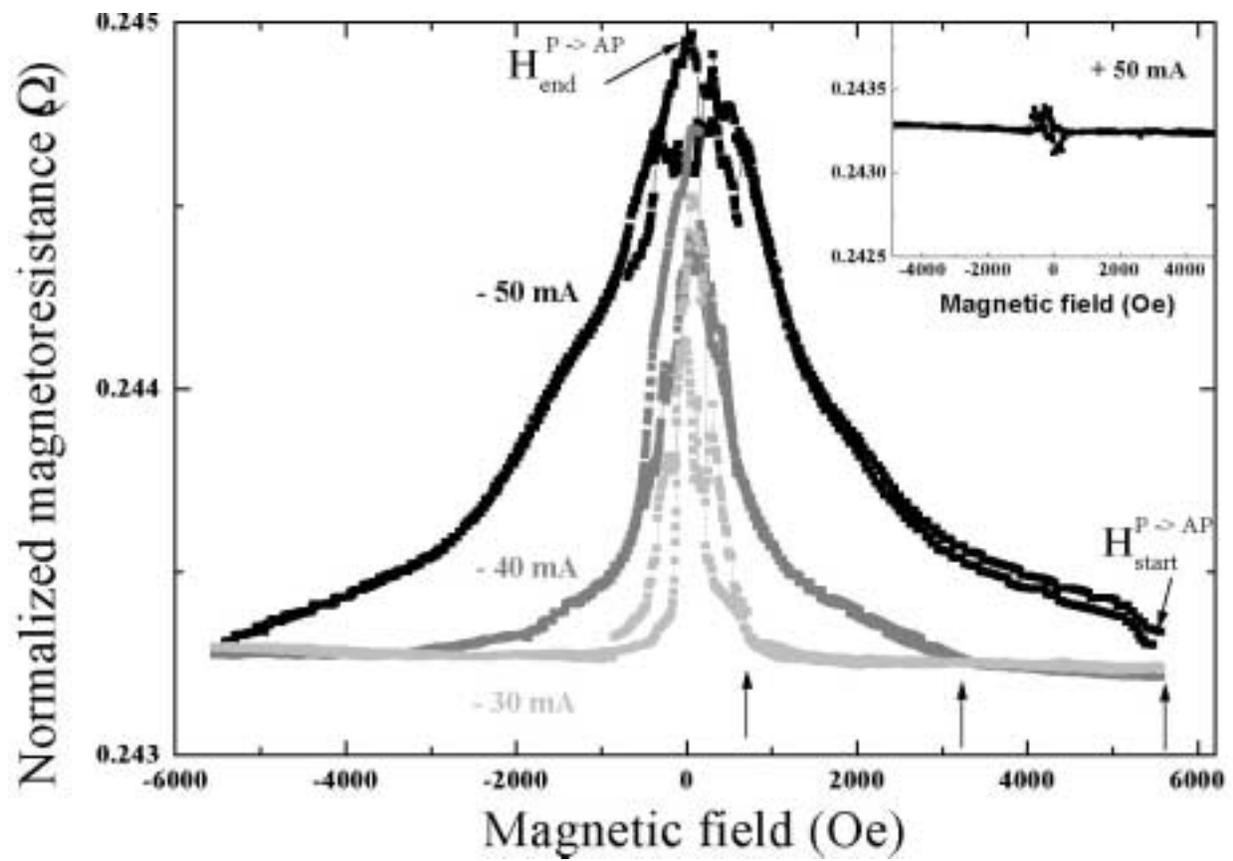

**Figure 4**

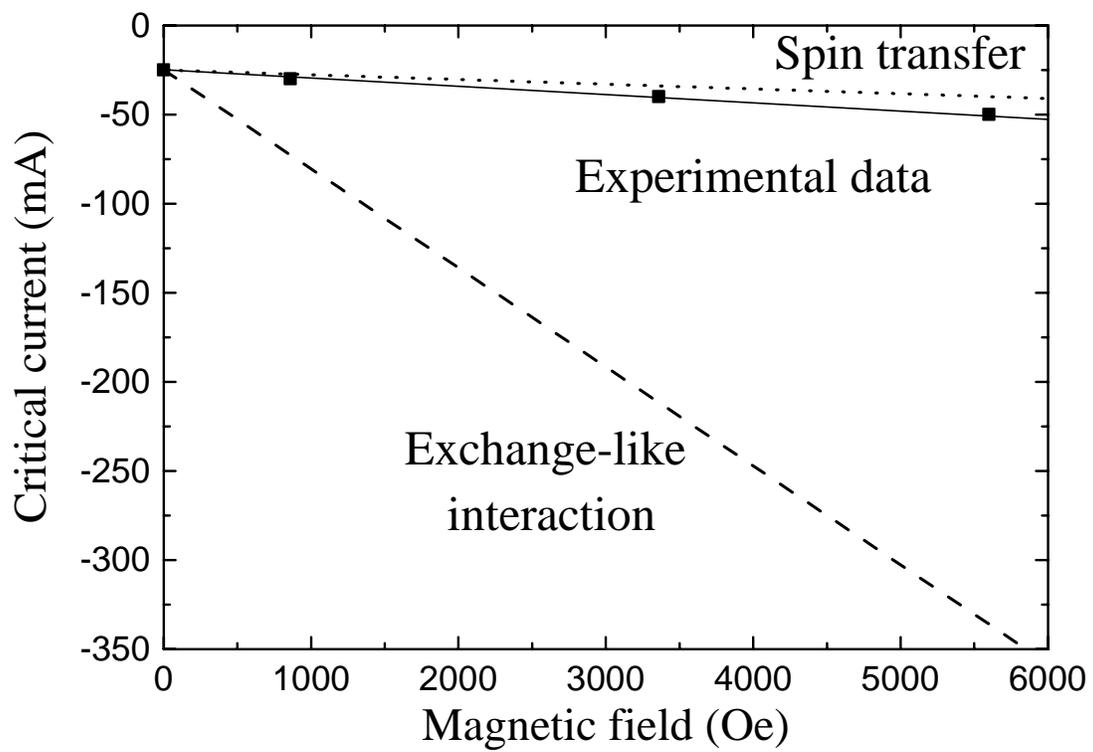

**Figure 5**[2]